\input harvmac
\input epsf
\Title{\vbox{\baselineskip12pt\hbox{EFI-96-07}
\hbox{hep-th/9603116}}}
{Vacua of M-theory and N=2 strings}

\centerline{David Kutasov, Emil Martinec,
and Martin O'Loughlin\footnote{*}{Supported
in part by Dept. of Energy grant DEFG02-90ER-40560.  }
}
\vskip .5cm
\centerline{\it Enrico Fermi Inst. and Dept. of Physics}
\centerline{\it University of Chicago}
\centerline{\it 5640 S. Ellis Ave., Chicago, IL 60637}
\vskip 1cm

\noindent
String and membrane dynamics may be unified into a theory
of 2+2 dimensional self-dual world-volumes living
in a 10+2 dimensional target space.
Some of the vacua of this M-theory are described by
the N=(2,1) heterotic string, whose target space theory 
describes the world-volume dynamics of 
2+2 dimensional `M-branes'. 
All classes of string and membrane theories are realized
as particular vacua of the N=(2,1) string:
Type IIA/B strings and supermembranes
arise in the standard moduli space
of toroidal compactifications, while
type ${\rm I}'$ and heterotic strings arise from a
$\bf Z_2$ orbifold of the N=2 algebra.
Yet another vacuum describes M-theory on a 
${\bf T}^5/{\bf Z}_2$ orientifold, 
the type I string on $ {\bf T}^4$,
and the six-dimensional self-dual string.
We find that open membranes carry `Chan-Paton fields'
on their boundaries, providing a common origin for 
gauge symmetries in M-theory.
The world-volume interactions of M-brane fluctuations
agree with those of Born-Infeld effective dynamics 
of the Dirichlet two-brane in the presence 
of a non-vanishing electromagnetic field on the brane.

\Date{3/96}
%\draftmode

\def\ie{{\it i.e.}}
\def\eg{{\it e.g.}}

\def\inbar{\,\vrule height1.5ex width.4pt depth0pt}
\def\IC{\relax\hbox{$\inbar\kern-.3em{\rm C}$}}
\def\IR{\relax{\rm I\kern-.18em R}}
\def\IP{\relax{\rm I\kern-.18em P}}
\def\Z{{\bf Z}}

\def\T{{\bf T}}
\def\S{{\bf S}}

%
%%%%%%%%%%%%%%%%%%%%%%%%%%%%%%%%%%%%
%
\def\np#1#2#3{Nucl. Phys. {\bf B#1} (#2) #3}
\def\pl#1#2#3{Phys. Lett. {\bf #1B} (#2) #3}
\def\prl#1#2#3{Phys. Rev. Lett. {\bf #1} (#2) #3}
\def\physrev#1#2#3{Phys. Rev. {\bf D#1} (#2) #3}

\def\jpa#1#2#3{J. Phys. {\bf A#1} (#2) #3}

\catcode`\@=11
\def\slash#1{\mathord{\mathpalette\c@ncel{#1}}}
\overfullrule=0pt

\def\FF{{\cal F}}

\def\II{{\cal I}}

\def\MM{{\cal M}}

\def\lam{\lambda}

\def\underrel#1\over#2{\mathrel{\mathop{\kern\z@#1}\limits_{#2}}}

\catcode`\@=12

%%%%%%%%%%%%%%%%%%%%%%%%%%%%%%%%%%%%%%%%%%%%%%%%%%%%%%%%%%%%%%

%
\def\det{{\rm det}}

\def\det{{\rm det}}
\def\exp{{\rm exp}}

\def\sst{\scriptscriptstyle}

\def\frac#1#2{{#1\over#2}}
\def\coeff#1#2{{\textstyle{#1\over #2}}}
\def\half{\frac12}
\def\hf{{\textstyle\half}}

\def\vev#1{\langle#1\rangle}
\def\d{\partial}

\def\oprime{{$\rm I'$}}

\def\j{{\bf j}}

\def\psibar{{\bar\psi}}
\def\phibar{{\bar\phi}}
\def\betabar{{\bar\beta}}
\def\phitilde{{\tilde\phi}}

\def\gammabar{{\bar\gamma}}

\def\dbar{{\bar \d}}

\def\Gbar{{\bar G}}

\def\Jbar{{\bar J}}

%%%%%%%%%%%%%%%%%%%%%%%%%%%%%%%%%%%%%%%%%%%%%%%%%%%%%%%%%%%%%%

\newsec{Introduction}

Strong/weak coupling duality relations among the diverse
supergravity theories, motivated by our
current understanding of string dynamics,
have begun to merge into a rather complete picture of
a single, unifying entity -- M-theory -- which reduces to
these various effective theories in different limits.
These limits include the various critical string theories --
type I, type IIA\&B, and heterotic -- as well as
eleven-dimensional supergravity.
It is clear from recent developments that there should exist
a formalism flexible enough to describe strings, membranes,
and perhaps higher $p$-branes in a unified way.
Such a formalism has been lacking until now, and thus so 
has a basic definition of M-theory; nevertheless,
the duality hypothesis has been used to infer
a number of its features (see for instance
\ref\sduality{C. Hull, hep-th/9512181.}
for a recent survey, and 
\nref\polchwit{J. Polchinski and E. Witten, \np{460}{1996}{525};
hep-th/9510169.}%
\nref\horwit{P. Ho\v rava and E. Witten, \np{460}{1996}{506};
hep-th/9510209.}%
\nref\witinst{E. Witten, \np{460}{1996}{541}; hep-th/9511030.}%
\nref\asselfdual{A. Strominger, hep-th/9512059.}%
\nref\townsend{P. Townsend, hep-th/9512062.}%
\nref\dm{K. Dasgupta and S. Mukhi, hep-th/9512196.}%
\nref\witfivebrane{E. Witten, hep-th/9512219.}%
\nref\schmidhuber{C. Schmidhuber, hep-th/9601003.}%
\nref\dmw{M. Duff, R. Minasian, and E. Witten, hep-th/9601036.}%
\nref\gimpolch{E. Gimon and J. Polchinski, hep-th/9601038.}%
\nref\ganhan{O. Ganor and A. Hanany, hep-th/9602120.}%
\nref\seibwit{N. Seiberg and E. Witten, hep-th/9603003.}
\refs{\polchwit-\seibwit}
for further results.).

Recently, two of us
\ref\km{D. Kutasov and E. Martinec, hep-th/9602049.}
showed that a route to the construction of M-theory
is provided by the N=(2,1) heterotic string, which 
describes the dynamics of M-theory
in particular backgrounds;
this leads for the first
time to a unified treatment of strings and membranes that
goes beyond their effective low-energy descriptions.
The central idea is that the
target space theory of the N=(2,1) heterotic string
generates the world volume dynamics of the various critical
string and membrane theories,
and hence serves as a sort of Rosetta stone to deciphering
the language of M-theory.
This target space theory is fundamentally 2+2 dimensional;
however, in the N=(2,1) string the worldsheet
gauge algebra includes a null current, which
restricts the kinematics to 1+1 or 2+1 dimensions.
This construction motivates a definition
of M-theory as a theory of 2+2 dimensional membranes
(we shall call them M-branes, following \townsend)
embedded in 10+2 dimensions
with a null reduction, whose dynamics is that of self-dual
dilaton gravity coupled to self-dual matter.
The choice of the null vector which defines the
gauged current determines whether one obtains
strings or membranes: Strings if the null vector is
entirely within the M-brane, membranes if it lies partly
in the M-brane world-volume and partly along
the transverse directions of the matter field-space.

The N=(2,1) string provides one with a tool to probe the
2+2 dimensional M-brane world-volume theory.
Ultimately, one will want to study directly
this 2+2 dimensional theory.
However, before attempting that,
it seems useful to get as much information 
as possible about aspects of the theory
described by N=(2,1) strings.
That is the goal of this paper.

The right-movers of the N=(2,1) string live in 2+2 dimensions,
the left-movers in 10+2 dimensions.
A common setting for the various backgrounds we
will consider is achieved
by compactifying all spatial coordinates of the target.
One finds a Narain moduli space of vacua $\MM$.
At generic points in $\MM$, the target space dynamics is essentially
1+2 dimensional, describing the world-volume
of a membrane.  A special alignment of the moduli gives the
1+1 dimensional dynamics of strings. 
Certain solitonic sectors of the theory appear to describe
$p$-branes.

The picture we find is consistent with string duality. 
We discuss three main classes of M-theory vacua.
Each class contains several seemingly different theories;
in our construction, they are continuously connected in
the moduli space of vacua $\MM$.
The three classes are:

\item{1)}
Type IIA/B strings
and the eleven-dimensional supermembrane.
For generic values of the moduli $\bf m\in\MM$,
one finds a stretched supermembrane; double dimensional
reduction yields the type IIA string,
while aligning the vector that defines the null reduction 
with the membrane world-volume yields the type IIB string.
This common setting for these three theories agrees with the fact
that type IIA/B strings are related by T-duality,
while the eleven-dimensional supermembrane describes the
strong coupling limit of the type IIA theory.

\item{2)}
The heterotic and type I strings, 
and M-theory on an $\S^1/\Z_2$ orientifold.
The generic point in moduli space describes 
an open supermembrane; 
different limits describe (a) the type \oprime\ string
with gauge group $SO(16)\times SO(16)$, and (b) the heterotic
string with gauge group $E_8\times E_8$.
Aligning the null reduction as above yields the
$SO(32)$ heterotic string.
Gauge groups for these theories appear
from a common source -- twisted
sectors of M-theory orientifolds;
our construction gives a new derivation 
of Chan-Paton factors, and clarifies the relation between
the sources of gauge groups for type I and heterotic strings.
The common setting for the above four theories
agrees with the fact that 
(i) $SO(32)$ and $E_8\times E_8$ strings,
and separately type I and type \oprime\ strings,
are related by T-duality; 
(ii) the heterotic and type I theories
are related by strong/weak coupling duality; and 
(iii) M-theory on $\S^1/\Z_2$ describes the strong coupling limit
of the $E_8\times E_8$ heterotic string.

\item{3)}
M-theory on a $\T^5/\Z_2$ orientifold, 
type IIA strings on a $\T^5/\Z_2$ orientifold,
type IIB strings on $K3$,
type I strings on $\T^5$, and 
the six-dimensional self-dual string
\ref\sds{E. Witten, contribution to STRINGS '95; hep-th/9507121.}.
The construction describes a membrane whose ends are trapped
on five-branes; different limits describe 
(a) a type IIA two-brane stretched between Dirichlet four-branes,
(b) the type IIA string on a $\T^5/\Z_2$ orientifold,
which is equivalent to the type I string
on $\T^5$, and 
(c) the self-dual string, which carries a `Chan-Paton' current algebra.
This common setting for the different theories agrees with the fact that 
(i) type IIA theory on a $\T^5/\Z_2$ orientifold is T-dual
to the type I theory on $\T^5$;
(ii) M-theory on a $\T^5/\Z_2$ orientifold describes certain
strong-coupling limits of type IIB strings on $K3$; and
(iii) the type IIB theory on $K3$ with an $A_1$ singularity
is equivalent to the eleven-dimensional M-theory with
coincident five-branes, both of which lead to a description
of the six-dimensional self-dual string.

\nref\bfk{W. Boucher, D. Friedan, and A. Kent, \pl{172}{1986}{316};
A. Schwimmer and N. Seiberg, \pl{184}{1987}{191}.}
\nref\lecht{S. Ketov, O. Lechtenfeld, and A. Parkes,
\physrev{51}{1995}{2872}, hep-th/9312150;
J. Bischoff, S. Ketov, and O. Lechtenfeld,
\np{438}{1995}{373}, hep-th/9406101.}
Let us now describe the plan of the present article.
In section two, we
review the relevant aspects of the N=2 right-moving and
N=1 left-moving gauge algebras of the
N=(2,1) string.  A new ingredient is the
twisted N=2 algebra \refs{\bfk, \lecht}
which is needed for some of the orbifolds in section four.
In section three, we describe the construction of the type IIB
string, and the eleven-dimensional/type IIA supermembrane
(item (1) above); here, the right-moving N=2 gauge
algebra is untwisted.  

Section four explores orbifolds/orientifolds of M-theory.
We begin with a construction of the type II string/eleven-dimensional
supermembrane on $K3$.
Then we consider the twisted N=2 algebra for the right-movers.
Using this algebra, we present a unified treatment of type I and 
heterotic theories (item (2) above).  
We also discuss the M-theory orientifold 
$\T^5/\Z_2$ (item (3) above).

In section five we resolve
a few apparent puzzles regarding our construction of the membrane;
the main observation is that there is naturally a non-zero
electromagnetic field on the world-volume of the membrane.
We summarize our results and discuss directions for future
research in section six.

%%%%%%%%%%%%%%%%%%%%%%%%%%%%%%%%%%%%%%%%%%%%%%%%%%%%%%%%%%%%%%

\nref\marcus{N. Marcus, 
talk at the Rome String Theory Workshop (1992); hep-th/9211059.}% 
\nref\ovthree{H. Ooguri and C. Vafa, \np{367}{1991}{83}.}%
\newsec{Review of N=(2,1) strings, and a twist}

The N=(2,1) heterotic string unites chiral N=2 local supersymmetry
for the right-moving degrees of freedom with antichiral N=1 local
supersymmetry for the left-moving degrees of freedom\foot{For more
on N=2 strings and references to earlier work see e.g. 
\refs{\marcus, \ovthree}. For more background on our construction see
\km.}.
Consistency also requires the introduction of a left-moving
anomaly-free U(1) super-current algebra \ovthree.
The right-moving gauge algebra is 
generated by the N=2 superconformal currents
$\bar T$, ${\bar G}_{\pm}$, $\bar J$.
There are two classes of boundary conditions allowed for
these currents \bfk.

First, since the supersymmetry currents $\Gbar^\pm$
are charged under the U(1) generated by $\Jbar$, one
may twist the former by a Wilson line of the latter.  This allows
for a continuous family of boundary conditions that interpolate
between the standard Ramond (R) and 
Neveu-Schwarz (NS) boundary conditions on the $\Gbar^\pm$.
Since the $U(1)$ symmetry generated by $\bar J$ is gauged, the
sectors corresponding to different boundary conditions are
identified.

Second, one can impose antiperiodic boundary conditions on
the U(1) current $\Jbar$, together with a $\Z_2$ twist interchanging
$\Gbar^+$ with $\Gbar^-$; this is usually called the twisted 
N=2 algebra \bfk. Such boundary conditions arise when
one gauges the above $\Z_2$ automorphism of the N=2
superconformal algebra, and lead to new sectors
of the Hilbert space.

In this section, we review these two classes of N=2 algebra,
together with the set of ghosts required for construction
of the BRST charge and corresponding representatives of physical states
for the right-movers
\ref\fms{D. Friedan, E. Martinec and
S. Shenker, \np{271}{1986}{93}.}, followed by a review of the
relevant aspects of the N=1
algebra of the left-movers.  In the next two sections
we will combine these two sectors,
first for vacua of the N=(2,1) string using the
untwisted N=2 algebra, and then for vacua using the
twisted N=2 algebra.

%%%%%%%%%%%%%%%%%%%%%%%%%%%%%%%%%%%%%%%%%%%%%%%%%%%%%%%%%%%%%%

\subsec{The right-moving N=2 algebra.}

The inequivalent representations of the N=2 algebra are characterized
by the boundary conditions on the currents.  One possibility
is to twist by U(1) spectral flow
\eqn\spflow{\eqalign{
  \bar T(2\pi)=&\bar T(0)\cr 
  \bar G^+(2\pi)=&-e^{2\pi iu}\bar G^+(0)\cr
  \bar G^-(2\pi)=&-e^{-2\pi iu}\bar G^-(0)\cr
  \bar J(2\pi)=&\bar J(0)\ .\cr
}}
For instance, what one usually calls the Neveu-Schwarz (NS) sector 
has $u=0$, and the Ramond (R) sector has $u=\half$.
In a theory with local $N=2$ supersymmetry
the U(1) current is gauged, and all such sectors are
equivalent, 
%\ref\ovtwo{H. Ooguri and C. Vafa, \np{361}{1991}{469}.}
corresponding to different Wilson lines of the gauge field.
To construct physical states in such a theory
one needs the ghosts for the
gauge algebra generated by  $\bar T$, ${\bar G}_{\pm}$, $\bar J$;
these will be denoted by $(\bar b,\bar c)$, 
$({\bar \beta}_\pm, {\bar \gamma}_\pm)$,
and $(\bar{\tilde b},\bar{\tilde c})$, respectively. In particular,
vertex operators depend on the fields
$\bar\phi_\pm$ arising from the bosonization of $\bar\beta_\pm$,
$\bar\gamma_\pm$ in the usual way \fms:
${\bar\beta}_\pm{\bar\gamma}_\pm=\dbar\bar\phi_\pm$; also
$\bar{\tilde b}\bar{\tilde c}=\dbar\bar{\tilde\phi}$. 

One may gauge in addition the $\Z_2$ symmetry taking
$\bar J\to -\bar J$ and exchanging $\bar G^\pm$. This gives rise
to a sector in which the $N=2$ algebra is twisted: 
\eqn\twisted{\eqalign{
  \bar T(2\pi)=&\bar T(0)\cr 
  \bar G_v(2\pi)=&\hf(\bar G^+ + \bar G^-)(2\pi)=-\bar G_v(0)\cr
  \bar G_a(2\pi)=&\hf(\bar G^+ - \bar G^-)(2\pi)=+\bar G_a(0)\cr
  \bar J(2\pi)=&-\bar J(0)\ .\cr
}}
In this case
it is convenient to use the corresponding combinations of the ghosts
which diagonalize the $\Z_2$ twist:
$\gammabar_v$, $\betabar_v$ and $\gammabar_a$, $\betabar_a$; the
corresponding bosonizations are $\betabar_v\gammabar_v=\dbar \phibar_v$,
$\betabar_a\gammabar_a=\dbar\phibar_a$.  

In flat spacetime $\IR^{2,2}$,
the right movers of the N=(2,1) heterotic
string are four real scalar fields $x^\mu$,
$\mu=0,1,2,3$ with signature $(-,-,+,+)$, and 
their superpartners under the $N=2$ superconformal
algebra, $\bar\psi^\mu$. 
The matter $N=2$ superconformal generators can be written as:
\eqn\rightNeqtwo{\eqalign{
  \bar T = & -\hf\dbar x\dbar x -\hf \psibar\dbar\psibar\cr
  \Gbar^\pm=&(\eta_{\mu\nu}\pm \II_{\mu\nu})\psibar^\mu\dbar x^\nu\cr
  \Jbar=&\II_{\mu\nu}\psibar^\mu\psibar^\nu\cr
}}
$\eta_{\mu\nu}$ is a metric on $\IR^{2,2}$, while the antisymmetric 
tensor $\II_{\mu\nu}$ corresponds to a non vanishing background field
on the world-volume of the brane. 
The magnitude of this field is determined
by the right-moving N=2 superconformal algebra; we will discuss
its role further in section 5.  
Throughout this paper we will use $\II_{03}=\II_{12}=1$.

The superconformal currents \twisted\ that diagonalize the $\Z_2$
of the twisted algebra are
\eqn\superconf{\eqalign{
  \Jbar=&\II_{\mu\nu}\psibar^\mu\psibar^\nu\cr
  \Gbar_v=&\eta_{\mu\nu}\psibar^\mu\dbar x^\nu\cr
  \Gbar_a=& \II_{\mu\nu}\psibar^\mu\dbar x^\nu.\cr
}}
Standard vertex operators for the untwisted algebra have the form:
\eqn\vvtwo{\eqalign{
V_{-1}(k)=&e^{-\bar\phi_+-\bar\phi_-}e^{ik\cdot x}\cr
V_0(k)=& \bar G^+_{-\half}\bar G^-_{-\half}e^{ik\cdot x}\cr}}
with the first line describing the vertex operator in the
$-1$ picture, and the second, in the $0$ picture.
To implement the twisted $N=2$ algebra \twisted\ one may
for example mod out by the $\Z_2$ symmetry $(x_1, x_3)\to-
(x_1, x_3)$, $(\bar\psi_1, \bar\psi_3)\to-(\bar\psi_1, \bar\psi_3)$,
taking $\bar J\to-\bar J$, $\bar G_v\to \bar G_v$, 
$\bar G_a\to -\bar G_a$ \superconf.
Standard vertex operators in the twisted sector have the form:
\eqn\twistverts{\eqalign{
  V^{\sst twisted}_{NS}=& e^{-\phibar_v-\half\phibar_a}\Sigma_{13} 
	S_{13}e^{\half\bar\phitilde}\; e^{ik\cdot x}\cr
  V^{\sst twisted}_{R}=& e^{-\half\phibar_v-\phibar_a}\Sigma_{13}
	S_{02}e^{\half\bar\phitilde}\; e^{ik\cdot x}\ ,\cr
}}
where $\Sigma_{13}$ and $S_{13}, S_{02}$ are the 
appropriate twist and spin fields 
that implement \twisted. The two operators 
$V_{NS}$, $V_R$ \twistverts\ are
identified by the gauged spectral flow.

%%%%%%%%%%%%%%%%%%%%%%%%%%%%%%%%%%%%%%%%%%%%%%%%%%%%%%%%%%%%%%%%%

\subsec{The left-moving $N=1$ sector.}

To describe the left-moving sector we need to find a $\hat c=10$ 
$N=1$ superconformal field theory (SCFT), to compensate the central
charge of the N=1 superconformal ghosts. The $2+2$ dimensional spacetime
coordinates $x^\mu$ (which are shared by the
left- and right-movers), and their superpartners $\psi^\mu$ form
a $\hat c=4$ SCFT.  
Naively, anomaly cancellation requires
an additional $\hat c=6$, but there
is a subtlety \ovthree. The $x^\mu$ live in 2+2 signature,
so the N=1 superconformal constraints are insufficient to
guarantee elimination of negative-norm states;
we need to gauge a null $U(1)$
supercurrent on the left side as well. The ghosts corresponding to this
$U(1)$ carry $\hat c=-2$ so overall we are looking for an internal
$\hat c=8$ ($c=12$) SCFT. A convenient representation is provided
by eight left-moving superfields $y^a+\theta\lambda^a$, $a=1, \cdots, 8$.
The $y^a$ are eight left-moving scalar fields living on the $E_8$
torus for modular invariance 
(more general possibilities exist and will be discussed below), 
while $\lambda^a$ are eight  Majorana -- Weyl fermions. 
The total $N=1$ superconformal algebra for 
the left movers can be taken to be:
\eqn\leftNeqone{\eqalign{
  T = &-\hf\d x\d x - \psi\d\psi
	-\hf\d y\d y -\lambda\d\lambda\cr
  G = &\psi\d x +\lambda\partial y.\cr
}}
As for the $N=2$ case, there are ghosts $b,c$, $\beta, \gamma$ 
needed to gauge \leftNeqone\ as well as ghosts of the gauged
$U(1)$ current algebra, $\tilde b,\tilde c$
and their superpartners $\tilde\beta, \tilde\gamma$.  
Some vertex operators
involve the bosonized ghost currents $\beta\gamma=\partial\phi$,
$\tilde b\tilde c=\d \tilde\phi$, $\tilde\beta
\tilde\gamma=\d\rho$.

A typical example of the left-moving part of a
physical NS vertex operator is
\eqn\nsvert{\eqalign{
  V_{-1}^a=&e^{-\phi}\lam^a e^{ik\cdot x}\cr
  V_0^a=&(\d y^a + ik\cdot\psi\lam^a)e^{ik\cdot x}\cr 
}}
Note that the $0$ picture vertex involves the translation
operator in the $y^a$ direction, $\d y^a$. It is natural to assume
that the scalar field $V^a$ lives on the
same torus as $y^a$. In particular, if we compactify $(x^i,
y^a)$, the target space theory lives on a spatial torus as well.
Orbifolds of the (2,1)
string that twist $x^i, y^a$ should correspond
to spacetime orbifolds.  We will use this interpretation
in sections three and four. 

The statistics of target space fields of the N=(2,1) string appears
to be determined solely by the NSR parity of the 
N=1 gauge algebra of the left-movers.
In constructions using only the untwisted N=2 algebra, one integrates
over Wilson line expectation values of a gauged U(1)
that interpolate between NS and R states; hence
target fermion parity cannot depend on right-moving
fermion boundary conditions.  Even in the twisted constructions,
instanton operators generate a discrete remnant of spectral
flow which flips this parity arbitrarily.
Thus we will assign target space statistics according to the
NSR parity of the left-movers alone.  It would be useful
to construct the target space
torus partition function in order to
verify that the appropriate signs arise from the different
sectors.

%%%%%%%%%%%%%%%%%%%%%%%%%%%%%%%%%%%%%%%%%%%%%%%%%%%%%%%%%%%%%%

\newsec{Vacua with maximal supersymmetry.}

\subsec{The general structure.}

To discuss different vacua of the N=(2,1) string in a unified
way, it is convenient to compactify
the spatial membrane world-volume coordinates $(x^2, x^3)$. 
We find ten spacelike right-moving scalars 
$z^i$, $i=2, \cdots, 11$, two
left-moving ones $\bar z^{\bar i}$, $\bar i=2,3$, and their 
superpartners, $\chi^i, \bar\chi^{\bar i}$:
\eqn\zchi{
\eqalign{
z^i=&(x^2, x^3, y^1, \cdots, y^8);\;\bar z^{\bar i}=(\bar x^2, \bar x^3)\cr
\chi^i=&(\psi^2, \psi^3, \lambda^1, \cdots, \lambda^8);\;
\bar \chi^{\bar i}=(\bar \psi^2, \bar \psi^3)\cr}}
Including the (non-compact) timelike superfields $(x^\alpha, \psi^\alpha)$,
$\alpha=0,1$ which may be denoted by $(z^\alpha, \chi^\alpha)$,
we have twelve left-moving fields $z^\mu$, $\mu=0,\cdots, 11$, and four
right-moving ones $\bar z^{\bar\mu}$, $\bar\mu=0,\cdots, 3$.
The left-moving superconformal algebra \leftNeqone\ is generated by:
\eqn\lll{T=-{1\over2}\partial z_\mu\partial z^\mu
-\chi_\mu\d\chi^\mu\ ;\;
G=\chi_\mu\partial z^\mu;\;\;\;\mu=0,\cdots 11\ .}
The $U(1)$ supercurrent we gauge has the general form:
\eqn\gaugedj{
\eqalign{
J=&\alpha\cdot\partial z;\;\;\; 
J|{\rm phys}\rangle=0\cr
\j=&\alpha\cdot \chi;\;\;\;\;\; 
\j|{\rm phys}\rangle=0\cr}}
$\alpha$ is an arbitrary null vector whose direction 
changes under $SO(2,10)$ Lorentz transformations.
For example, we may take $\alpha$ to point
in the $(x^1,x^3)$ direction, so that $J=\partial x^1+\partial x^3$,
$\j=\psi^1+\psi^3$. 

The momenta of the compact scalars $(z^i, \bar z^{\bar i})$,
$i=2,\cdots, 11$, $\bar i=2,3$, live on an even self-dual 
Lorentzian lattice\foot{A word of caution: one should not confuse
the two Lorentzian spaces with signature (10,2) discussed here.
The left movers of the N=(2,1) string live in a 2+10 dimensional
target space parametrized by $z^\mu$; separately, if we compactify
all ten spatial dimensions $(z^i, \bar z^{\bar i})$ \zchi, 
their left- and right-moving momenta live on a (10,2) dimensional
Narain lattice $\Gamma^{10,2}$.}
$\Gamma^{10,2}$. The moduli space of vacua, $\MM$ 
is locally $\MM\simeq SO(10, 2)/SO(10)\times SO(2)$. 
Here we want to study some of its properties. 

The theory is supersymmetric in target space. The supercharges form a $16$ 
of $SO(1,9)$. They can be constructed in the standard fashion \fms\
out of the $SO(2,10)$ spin fields $S_\alpha$ for the $12$ fermions 
$\chi^\mu$, and the bosonized ghost current $\phi$. Since one is twisting
the fermionic current $\j$ \gaugedj, the supercharges contain the 
dimension $-1/8$ twist field $\exp(\rho/2)$ for
the bosonic ghosts $\tilde \beta, \tilde\gamma$:
\eqn\sptm{Q_\alpha=\oint dz e^{-{\phi\over2}+{\rho\over2}}
S_\alpha.}
The $SO(2,10)$ spin field $S_\alpha$ has $32$ components, but only $16$
of those satisfy the second gauge condition in \gaugedj. These are 
supercharges satisfying 
\eqn\condtn{\gamma^1\gamma^3Q=Q}
which, taken together with the twelve-dimensional chirality projection
\eqn\tcondtn{\left(\prod_{\mu=0}^{11}\gamma^\mu\right)Q=Q\ ,}
implies that the physical supercharges form a Majorana-Weyl
spinor of the $SO(1,9)$ acting on $(x^0, x^2, y^1, \cdots, y^8)$.

The physical states assemble naturally into a gauge field in $2+10$ 
dimensions: 
\eqn\scalrs{V=e^{-\phi}\xi_\mu\chi^\mu e^{-\bar\phi_+-\bar\phi_-}
e^{ik\cdot z+i\bar k\cdot\bar z}.}
The polarization and momentum vectors satisfy the familiar physical state
conditions:
\eqn\phst{
\eqalign{
k^2=&\bar k^2=0\cr
k\cdot\xi=&0;\;\;\xi\sim \xi+\epsilon k.\cr}}
The first line states that $V$ is massless, while the
second is a consequence of gauge invariance in $2+10$
dimensions. 

In addition to \phst, the gauged supercurrent \gaugedj\ leads to the 
following constraints on the physical states \scalrs:
\eqn\addcons{k^3+k^1=0;\;\;\;\xi^3+\xi^1=0.}
As discussed in \km, this gauge invariance effectively sets 
$k^3=k^1=\xi^3=\xi^1=0$, removing two target space dimensions.
Thus, to count physical degrees of freedom contained in 
\scalrs, we start with twelve polarizations of the $2+10$ 
dimensional gauge field $V$; the null gauging reduces it to a gauge field 
in $1+9$ dimensions; gauge invariance \phst\ implies that there
are eight physical degrees of freedom.

Supersymmetry \sptm\ pairs these eight scalars with
eight physical fermionic degrees of freedom:
\eqn\frms{F=u^{\bar\alpha}(k) e^{-{\phi\over2}+{\rho\over2}}S_{\bar\alpha}
e^{-\bar\phi_+-\bar\phi_-} e^{ik\cdot z+i\bar k\cdot\bar z}.}
The polarization vector $u$ satisfies a Dirac equation, and a super -- 
$U(1)$ constraint analogous to \condtn:
\eqn\Feq{k_\mu\gamma^\mu u=0;\;\;\; \gamma^1\gamma^3 u=u.}
The two together describe eight propagating degrees of freedom, in 
agreement with supersymmetry.

The physical states assemble naturally into a $1+9$ dimensional
vector superfield (after imposing the null reduction), but the
theory does not in general live in $1+9$ dimensions. The spectrum
of operators \scalrs, \frms, and supersymmetry structure \sptm\
is identical to that arising in the D-brane construction of 
\ref\polch{J. Polchinski, \prl{75}{1995}{4724}; hep-th/9510017.}.
By analogy, one would expect  to find different sectors 
of the theory in which the momenta
of $V,F$ are $1+p$ dimensional, and their interactions describe 
$p$-brane world-volume dynamics. In the N=(2,1) construction,
strings ($p=1$) and membranes
($p=2$) are singled out by the decompactification limits of
$\Gamma^{10,2}$; if we go to the boundary of moduli space $\MM$ 
where the theory describes 
two non-compact scalar fields and eight left-moving
scalars living on the $E_8$ torus, one finds \km\ that the target
space of the $N=(2,1)$ string is either $1+1$ or $1+2$ dimensional,
depending on the orientation of the null vector \gaugedj.
We will describe here the string and membrane constructions, 
noting only that higher $p$-branes originate in mixed momentum/winding
sectors of $\Gamma^{10,2}$; we leave
a detailed analysis of these higher $p$-branes to future work.

\subsec{Type II strings and the supermembrane.}

An arbitrary Lorentzian even self-dual lattice $\Gamma^{10,2}$ 
can be brought by an $SO(10,2)$ Lorentz transformation to a
standard form
\eqn\ff{\Gamma^{10,2}\simeq \Gamma^{1,1}\times\Gamma^{1,1}
\times \Gamma^8}
with the two $\Gamma^{1,1}$ factors describing two
compact scalar fields, say $x^2, x^3$ and $\Gamma^8$ describing
the remaining left-moving scalars, $y^1, \cdots, y^8$ living on the
$E_8$ torus.
Different vacua of the N=(2,1) string can be studied by fixing 
the $\Gamma^{10,2}$ \ff, and varying the direction of the null
vector $\alpha$ \gaugedj\ with respect to it. 

We can choose the timelike component of $\alpha$ to point in the
$x^1$ direction; this defines $x^1$.
There are three logical possibilities for the
spacelike component of $\alpha$ which we consider
in turn: it could (a) lie purely in the 
$\Gamma^{1,1}\times\Gamma^{1,1}$ part of \ff,
(b) lie purely in $\Gamma^8$, or (c) have components in both.

In the first case, the null gauging effectively removes
one of the two circles, and we are left with momenta
lying in $\Gamma^{9,1}\simeq \Gamma^{1,1}\times\Gamma^8$.
For example, we may take $J=\partial x^1+\partial x^3$, 
in which case the $\Gamma^{1,1}$ momenta $(k_2, \bar k_2)$ 
describe  the compact scalar field $x^2$, and the momentum 
in the $\vec y$ direction $\vec p$ belongs to the $E_8$ root
lattice, $\Gamma^8$. We will focus on momentum modes of the
compact scalar $x^2$. As discussed in \km, winding modes
describe another copy of the structure (in this case a ``winding
string''), while mixed momentum -- winding sectors give rise
to higher $p$-branes, and will not be discussed here in any detail.

The zero winding number sector on $\Gamma^{1,1}$ together
with the timelike non-compact dimension $x^0$ describes a
1+1 dimensional target space $(x^0, x^2)$. 
The physical spectrum includes a ten-dimensional $U(1)$
gauge superfield $V$, $F$ \scalrs, \frms, dimensionally reduced
to two dimensions. This gives rise to an $(8,8)$ supersymmetric
theory of eight scalars and fermions (the transverse
components of the gauge field). In \km\ this target space
theory has been interpreted as the world sheet theory describing
a type IIB string in static gauge\foot{The fact that this
is the type IIB string (as opposed to a type IIA one) follows
from the structure of the supersymmetry algebra, and in particular
from the fact that left- and right-moving fermions $F$ \frms\
belong to different spinor representations of ${\rm Spin}(8)$.}. 
The transverse dimensions of that string
are compactified on an eight-torus. The spatial world sheet 
coordinate $x^2$ is identified with the longitudinal direction
on the string, hence the latter is compact as well. Thus we naturally
find the type IIB string compactified on a spatial torus $\T^9$, with
only time, which is identified with the world sheet time 
coordinate $x^0$ by the static gauge condition, remaining non-compact. 

Consider next the case where the null vector $\alpha$ has a 
spatial component purely in $\Gamma^8$, 
say $J=\partial x^1+\partial y^1$. The zero winding sector
in the two spatial circles describes this time a 1+2 dimensional
target space  parametrized by $(x^0, x^2, x^3)$, describing \km\
the world-volume of a supermembrane. Dimensional
reduction of \scalrs, \frms\ gives rise now to seven scalars and
a gauge field on the world-volume, as well as eight fermionic states,
related by N=8 Green-Schwarz supersymmetry on the world-volume.

Type IIA strings appear in this formalism by double
dimensional reduction of the membrane construction.
It is well known that dimensional reduction of
eleven-dimensional supergravity gives the non chiral type IIA
supergravity in ten dimensions. Thus, the type IIA string can be obtained
by taking the radius of (say) $x^3$ to zero in the second construction
described above\foot{Actually the limit $R_3\to0$ yields a type IIB
string propagating in a spacetime one of whose spatial dimensions is a 
circle with vanishing radius. T-duality relates that to a type IIA
string on a large circle.}.  

Note that we always find sixteen supercharges on the world-volume of
the brane, half of the number of supercharges in a covariant
description of the theory. The reduction is due to the imposition
of static gauge, which allows only supercharges that close on translations
along the world-volume of the brane.

The case when the spatial part of the null vector $\alpha$ 
has components both in $\Gamma^8$ and in $\Gamma^{1,1}\times
\Gamma^{1,1}$ is also interesting, as it allows one to 
continuously go from the membrane to the string vacuum of
M-theory.
Indeed, suppose:
\eqn\jjjj{J=\partial x^1+ a\partial x^3 + b\partial y^1}
with $0\le a,b \le 1$, and $a^2+b^2=1$.
As we saw before, for $a=0$ \jjjj\ describes a 
supermembrane world-volume,
while for $b=0$ it describes a type IIB string world sheet.
If $(x^2, x^3)$ live on circles of radii $(R_2, R_3)$, for general
$a,b$ \jjjj\ the target space is 1+2 dimensional with the
spatial dimensions effectively living on circles of radii 
$(R_2, bR_3)$; thus, indeed in the limit $b\to0$ the width of
the membrane goes to zero and it turns into a string.

To summarize, in this section we have shown that the moduli space
of the simplest N=(2,1) string vacua describes in target space
the world-volume dynamics of type IIA/B strings and the 
eleven-dimensional supermembrane. All theories naturally arise compactified
on a spatial torus. Which description is most natural changes as we
vary the moduli. 

The fact that these vacua of M-theory are closely related is known
\nref\witsdvd{E. Witten, \np{443}{1995}{85}; hep-th/9503124.}%
\nref\aspinwall{P. Aspinwall, hep-th/9508154.}%
\refs{\witsdvd,\aspinwall}. 
Compactified type IIA and IIB strings  are related by T-duality,
while the eleven-dimensional supermembrane and IIA theories are
related by strong-weak coupling ``duality'' (the former is the strong
coupling limit of the latter).

%%%%%%%%%%%%%%%%%%%%%%%%%%%%%%%%%%%%%%%%%%%%%%%%%%%%%%%%%%%%%%

\newsec{Orbifolds/orientifolds of M-theory.}

In section 3 we have studied some vacua of M-theory with the
largest possible supersymmetry.  This excludes many interesting
classes of vacua, such as type I and heterotic theories on tori,
as well as type II theories and supermembranes 
on K3 manifolds.  In this section
we will examine some such vacua; the discussion is not intended
to be exhaustive -- the purpose is to demonstrate that many vacua
of M-theory with reduced supersymmetry are realized by the N=(2,1)
string.  Another motivation is to compare the structures emerging
from our version of M-theory to predictions of string duality 
that exist in the literature. 
We find qualitative agreement with those predictions.

\subsec{Type II strings and M-theory on K3.}

In section 3 we have discussed a vacuum of the N=(2,1) string
describing the world-sheet of a type IIB string compactified
on a nine-torus. It is not difficult to describe a type IIB 
string propagating on $K3\times \T^5\times \IR$.
Recall that the type IIB theory arose when we gauged $J=\partial 
x^1+\partial x^3$; the target space of the N=(2,1) string 
(identified with the world sheet of a type IIB string) was
the 1+1 dimensional space parametrized by $(x^0, x^2)$.
To study the theory on a $K3$ orbifold $\T^4/\Z_2$ 
we can orbifold the N=(2,1) string by the discrete symmetry
\eqn\tiib{\eqalign{
(y^5, y^6, y^7, y^8) &\rightarrow - (y^5, y^6, y^7, y^8)\ ,\cr
(\lam^5, \lam^6, \lam^7, \lam^8) &
	\rightarrow - (\lam^5, \lam^6, \lam^7, \lam^8)\ ,
}}
inverting four of the left-moving spatial coordinates. 
As explained in section 2, the world sheet fields $y^a$
are essentially identified with the spacetime coordinates,
so this orbifold can in fact be thought of as acting in the
1+9 dimensional spacetime in which the type IIB string lives\foot{
Indeed, the physical vertex operators \scalrs\ 
$V^a  = e^{-\phi} \lambda^a  e^{-{\phibar}_+ -{\phibar}_-} 
e^{i k\cdot x}$ with $a=5,6,7,8$ are inverted under the action of \tiib\ as
appropriate for a spacetime orbifold.}.

Gauging the symmetry \tiib\ breaks the $SO(1,9)$ Lorentz symmetry
of the vacuum to $SO(1,3)\times SO(6)$. The spacetime supercharges
\sptm\ constructed in section 3 decompose as $(2,4)\oplus(\bar 2,
\bar 4)$ under the unbroken Lorentz group. Half of them, say the $(2,4)$,
survive the projection \tiib. The unbroken supercharges form
a (2,0) supersymmetry algebra in the six dimensions orthogonal to $K3$. 
This chiral supersymmetry is precisely what one expects for a type 
IIB string on $K3$ in static gauge.

To find physical states in the untwisted sector 
one needs to impose invariance 
under the $\Z_2$ symmetry \tiib\ on the states \scalrs,
\frms\ discussed in section 3.
This eliminates four of the eight scalars $V$ and their superpartners;
the invariant states are the ones describing translations in the
$y^1, \cdots, y^4$ directions,
\eqn\tiic{
V^a  = e^{-\phi} \lambda^a  e^{-{\phibar}_+ -{\phibar}_-} 
e^{i k\cdot x} ;\;\; a = 1,\cdots ,4.}
In addition, there are four massless scalars
(and their superpartners)  coming from the twisted sector:
\eqn\tiid{ 
W^{i\alpha}  = e^{-\phi} \Sigma^i S^\alpha\  e^{-{\phibar_+} - 
{\phibar_-}}\ e^{ik\cdot x}\ .}
Here $\Sigma^i$ are twist fields for 
the chiral scalars $y^5,\cdots ,y^8$; $S^\alpha$
are spin fields constructed out of $\lambda^1, \cdots, \lambda^4$.
On a $\T^4/\Z_2$ there are sixteen fixed points, but the $y^a$ are
chiral scalars, so there are four twist fields $\Sigma^i$ (which can be 
described as spin fields after fermionization of $y^a$).
The spin fields $S^\alpha$ belong to a $2$ of $SO(4)$; their chirality
is fixed by the GSO projection (or equivalently 
by mutual locality with the unbroken supercharges).
Finally, out of the four chiral twist fields $\Sigma^i$
only two survive the GSO projection. Thus, \tiid\ describes four
scalar fields in the two-dimensional target space
$(x^0, x^2)$; they can be thought of
as Nambu-Goldstone modes for the translation invariance%
\foot{Indeed, they transform as $2\times\bar 2=4$
under the $SO(4)$ acting on $\T^4/\Z_2$.} 
broken by the orbifold \tiib.
The spectrum is in agreement with what one expects;
the scalars $V^a$, $W^{i\alpha}$ are the transverse coordinates
of the type IIB string, living in $\T^4$, $\T^4/\Z_2$ respectively.

One can repeat the analysis for the eleven-dimensional
vacuum of M-theory. Gauging, as in section 3, $J=\partial x^1+
\partial y^1$, we find in the target space of the N=(2,1) string
a membrane world-volume $(x^0, x^2, x^3)$. 
Twisting by \tiib\
leads to the eleven-dimensional theory on $K3$. 
The supersymmetry and spectrum are similar 
to those discussed above; the only difference is that one of the scalar
fields \tiic, $V^1$ is eliminated by the gauge conditions and is replaced
by a three-dimensional gauge field (which in three dimensions 
is equivalent to a scalar ). Double dimensional reduction
of the supermembrane theory on $K3$ leads to a type IIA string, as before.

According to string duality, type IIB strings on $K3$ have a large
duality group $SO(21, 5; \Z)$ \refs{\witsdvd,\sduality}. 
This symmetry would only become
apparent in our formalism after second quantization of the N=(2,1)
string, and it is not surprising that we find a unique theory 
in the restricted class of vacua described
by the N=(2,1) string. 
It has also been argued \refs{\dm, \witfivebrane} that a compactification
of the eleven-dimensional M-theory on $\T^5/\Z_2$ is equivalent
to the type IIB string on $K3$, and so in the next 
subsection we turn to a set of
constructions that will allow us to study that equivalence. 

String duality also suggests that the eleven-dimensional
theory and the type IIA string compactified on $K3$ are
equivalent to the heterotic string compactified on a 
three- and four-torus, respectively. 
That duality plays an important role in
analyzing the strong coupling dynamics of heterotic and type II 
strings in dimensions four through seven. The IIA -- heterotic duality
will make an appearance in the constructions of the next 
subsection.

\subsec{Twisted N=(2,1) strings and open membranes.}

In this subsection we will discuss a class of vacua of M-theory
in which, in addition to N=(2,1) chiral world-sheet supergravity,
we gauge the $\Z_2$ automorphism 
\twisted\ of the N=2 superconformal algebra
discussed in section 2. This leads to M-theory on various
orientifolds.  Orientifolds arise rather generally in our construction
since the N=(2,1) string target space describes (stretched)
strings and membranes in static gauge. Thus, some coordinates
play a double role as both M-brane 
world-volume and spacetime coordinates.
Twisting these coordinates leads to 
combined world-sheet/spacetime
orbifolds or orientifolds (as well as $p$-branes).  

We start with a description of M-theory on an
$\S^1/\Z_2$ orientifold, and relate it to
heterotic and type I string theories.
Then we move on to a more intricate example:
M-theory on a $\T^5/\Z_2$ orientifold, 
which is expected \refs{\dm,\witfivebrane}
to describe certain strong coupling limits of type IIB strings 
on $K3$, as well as type I (and heterotic)
strings on $\T^4$.
In all cases, the relations between the different vacua we find 
agree with expectations from string duality.

\medskip

\noindent{}{\it a) Type I and heterotic strings.}

Type I and heterotic strings with gauge group $SO(32)$
are believed to be related by strong-weak coupling duality
\refs{\witsdvd,\polchwit}.
In fewer than ten non-compact dimensions
they are furthermore related to the $E_8\times E_8$
heterotic string by heterotic T-duality. The latter
is in turn described at strong coupling by a certain $\Z_2$ 
orientifold of the eleven-dimensional version of M-theory \horwit.
In this subsection we will construct a class of N=(2,1)
string vacua which makes these relations manifest.
In different regions of the moduli space of N=(2,1) string vacua
the theory we will construct looks like:

\item{1.} M-theory on $\S^1/\Z_2$ 
(or equivalently the $E_8\times E_8$
heterotic string).

\item{2.} The $SO(32)$ heterotic string (with gauge group possibly
broken to $SO(16)\times SO(16)$ by Wilson lines).

\item{3.} The type I$^\prime$ theory with gauge group 
$SO(16)\times SO(16)$ (which is T-dual to
a type I theory with the same gauge group).

In the spirit of the discussion of section 3, we compactify the spatial 
coordinates $(z^i, \bar z^{\bar i})$ \zchi\ on $\Gamma^{1,1}\times 
\Gamma^{1,1}\times \Gamma^8$ \ff. Then twist by:
\eqn\hetonetwist{\eqalign{
  (x^1,x^3,y^1,\ldots,y^8)&\rightarrow -(x^1,x^3,y^1,\ldots,y^8)\ ;\cr
  (\psi^1,\psi^3,\lam^1,\ldots,\lam^8)&
	\rightarrow -(\psi^1,\psi^3,\lam^1,\ldots,\lam^8)\ ;\cr
  (\psibar^1,\psibar^3)&\rightarrow -(\psibar^1,\psibar^3)\ ;\cr
}}
Note that the $\Z_2$ symmetry \hetonetwist\ twists the left- and right-moving
gauge algebras in the manner described in section 2:
\eqn\hetonealg{\eqalign{
  \Gbar^+\longleftrightarrow\Gbar^-\ ;&\quad \Jbar\rightarrow -\Jbar\cr
	G\rightarrow G\ ;&\quad J\rightarrow -J\cr
}}
Thus, in the twisted sector of the orbifold the $N=2$ algebra
will be twisted. 
Also, \hetonetwist\ describes an asymmetric orbifold. The fields
$y^1,\cdots, y^8$ are chiral, while $x^1, x^3$ have both left-
and right-moving components. Correspondingly, the orbifold
\hetonetwist\ is chiral for the $y^i$ and non-chiral for $x^1, x^3$.

The timelike component of the gauged current $J$ \gaugedj\
can be chosen to be $\partial x^1$, without loss of generality.
To specify the vacuum, one needs to fix
the orientation of the spatial component
of $J$. Some general properties 
of the resulting vacua are independent of that choice and we
will discuss these first.  

Half of the supercharges \sptm\ are broken by the orbifold
\hetonetwist. The surviving supercharges satisfy:
\eqn\orbsup{\gamma^0\gamma^2Q=Q.}
The physical states fall into two classes. In the untwisted
sector we have the eight scalar modes \scalrs, and their superpartners
under the unbroken supersymmetries \orbsup.
The twisted sector gives rise to sixteen fermionic states
from each of the two fixed points of the (non-chiral) $x^3$ orbifold:
\eqn\feri{F^i=e^{-\half\phi}\sigma\Sigma^i
e^{-\phibar_v-\half\phibar_a}\ e^{ik\cdot x}}
where $\Sigma^i$, $i=1, \cdots, 16$ are the chiral
twist fields for $(y^1, \cdots, y^8)$, and $\sigma$ contains
the necessary twist and spin fields for $x^\mu$, $\psi^\mu$,
$\bar\psi^\mu$ and the $U(1)$ ghosts:
\eqn\newsgma{\sigma=\sigma_{13}S_{02}\bar S_{13}
e^{\half\tilde\phi+\half\tilde\phibar}.}
The detailed structure of the theory
depends on the particular construction. We next describe the different
possibilities. As for other constructions encountered in this paper,
one can pass between the various theories we find
by changing the Narain moduli of the N=(2,1) string.

To get M-theory on $\S^1/\Z_2$ we choose the gauged $U(1)$
current to be:
\eqn\mth{J=\partial x^1+\partial y^1.}
Before gauging \hetonetwist\ this gives
rise to the eleven-dimensional supermembrane theory
with world-volume $(x^0, x^2, x^3)$ described in section 3.
The orbifold introduces two boundaries on the world-volume,
at $x^3=0,\pi R_3$, and correspondingly two boundaries
in spacetime in the direction identified
with $x^3$ by the static gauge condition.

To analyze the spectrum we are instructed to first find the
states in the untwisted sector \scalrs, \frms\ that are invariant under
\hetonetwist. The physical polarizations of the bosons $V$ \scalrs\
that survive the projection are:
\eqn\vaa{V^a=e^{-\phi}\lambda^ae^{-\phibar_a-\phibar_v}
e^{ik\cdot x}\cos k_3 x^3;\;\;\;a=2,\cdots, 8}
and the three-dimensional gauge field
\eqn\gff{\xi_\mu A^\mu=e^{-\phi-\phibar_a-\phibar_v}e^{ik\cdot x}
\left(\xi_0\psi^0\sin k_3x^3+\xi_2\psi^2\sin k_3 x^3+
\xi_3\psi^3\cos k_3x^3\right).}
In the above two equations $k$ denotes the momentum in the 
$(x^0, x^2)$ directions; $\xi$ is the polarization of the gauge field.
Dualizing the gauge field $A^\mu$ via $\epsilon_{\mu\nu\lambda}F^{\mu\nu}
=\partial_\lambda\Phi$ we find  \vaa, \gff\ that
both $V^a$ and $\Phi$ satisfy Neumann boundary conditions
at $x^3=0, \pi R_3$. Thus the theory contains eight fluctuating fields
both in the bulk of the membrane and on the boundary\foot{In particular,
\hetonetwist\ describes an $\S^1/\Z_2$ orientifold and not
a $\T^9/\Z_2$ one as one might have thought.}. 
It describes a membrane of the eleven-dimensional
theory stretched between two nine-branes
with world-volumes parametrized by $\{x^0,x^2,\Phi, V^a\}$; 
in the original presentation, one has what would be
described in type IIA language as a Dirichlet two-brane stretched
between Dirichlet eight-branes (see figure 1).
Note, however that there is a basic distinction between our
description of the membrane and that of type IIA theory 
\ref\ortfld{J. Polchinski, S. Chaudhuri and C. Johnson, hep-th/9602052.}.
In that theory the IIA string is taken to be fundamental,
and the two-brane is treated as a soliton; 
our theory treats the two-brane as fundamental.
The unbroken supersymmetry \orbsup\ pairs
the scalars with eight fermions $F$ \frms.
It is not obvious at this level what space the bosonic physical 
excitations $(V^a, \Phi)$ live on. We believe that they 
live on an eight-torus, 
but a definite answer must await a more detailed analysis.

Next we turn to the twisted sectors of the orbifold.
On each of the two-dimensional boundaries of the world-volume,
which are labeled by $(x^0, x^2)$, we find sixteen chiral fermions
\feri\ that are inert under the unbroken supersymmetry. 
In the limit $R_3\to
0$, the N=(2,1) string target space becomes two-dimensional; 
the fields living on the world-volume are eight massless scalars
$V^a, \Phi$ parametrizing $\T^8$, their eight right-moving
superpartners $F^\alpha$, and thirty-two left-moving fermions
$F^i$, sixteen from each of the fixed points. The charges
\orbsup\ form an $(8,0)$ supersymmetry algebra, 
$\{Q_\alpha, Q_\beta\}=\delta_{\alpha\beta}(p^0+p^2)$.
This is the right structure for a heterotic string compactified
on $\T^8$. For finite $R_3$ we find an open membrane with
finite extent. This is precisely the structure 
expected for the $E_8\times E_8$ string at finite coupling
\horwit. Our analysis provides new evidence for the 
$E_8$ current algebras conjectured in \horwit\ to live
on the boundaries of the membrane world-volume.

\vskip 1cm
{\vbox{{\epsfxsize=1.5in
    \centerline{\epsfbox{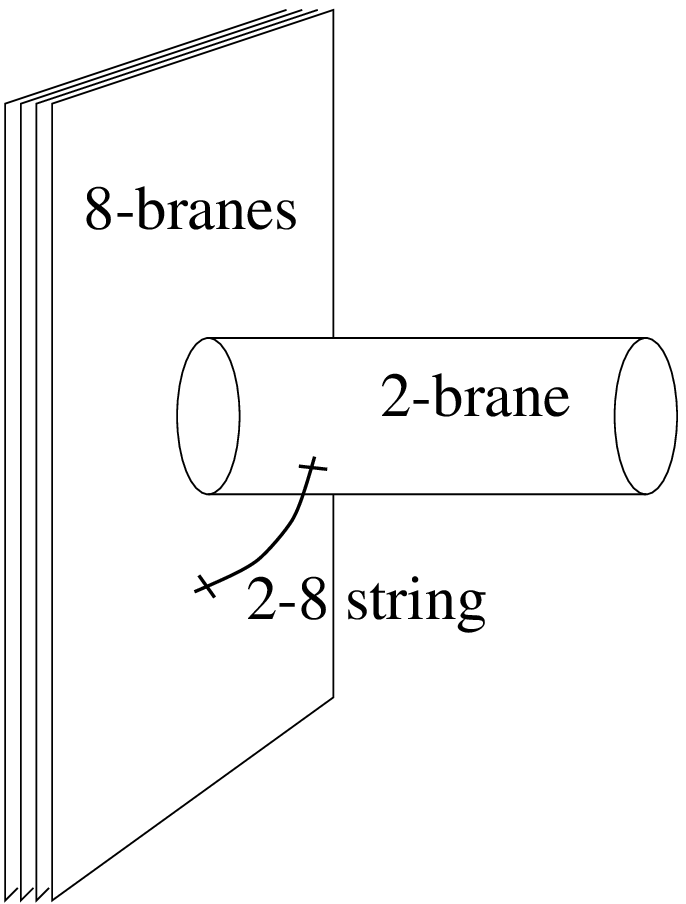}}
\vskip .5cm
    {\raggedright\it \vbox{
{\bf Figure 1.}
{\it A Dirichlet two-brane ends on a stack of Dirichlet eight-branes;
2-8 strings generate `Chan-Paton' dynamics on the two-brane boundary.
These degrees of freedom are the universal source of fundamental gauge
symmetry in M-theory.}
 }}}}
    \bigskip}

The $SO(32)$ heterotic string vacuum of the theory is obtained
by gauging 
\eqn\mthtwo{J=\partial x^1+\partial x^3.}
This gives a theory living in a 1+1 dimensional target 
space parametrized by $(x^0, x^2)$. The spectrum and
supersymmetry algebra \orbsup, \feri\ are again those of the
heterotic string. The fermions $F^i$ still split into
two groups of sixteen coming from each of the two fixed
points of the $x^3$ orbifold, but now because of the null gauging
\mthtwo\ there is nothing in the theory that can sense the
separation in the $x^3$ direction. Thus, the thirty two fermions
$F$ appear symmetrically, just as one would expect for an $SO(32)$
heterotic string.  It is nevertheless possible that a non-perturbative
analysis of the N=(2,1) string would reveal that $SO(32)$
is broken by a Wilson line to $SO(16)\times SO(16)$. 
We leave this question to future work.

The type \oprime\ string is in our formulation
a close relative of the $E_8\times E_8$ heterotic string, corresponding 
to a different dimensional reduction of the membrane discussed above.
Gauging \mth\ again we get a membrane with world-volume
$(x^0, x^2, x^3)$. To get the weakly coupled $E_8\times
E_8$ heterotic string we took the radius $R_3$ to zero. 
If instead we consider the limit $R_2\to0$ we find a theory 
with target space $(x^0, x^3)$; $x^3$ lives on a line
segment. The target space theory describes a type 
IIA string stretched between two eight-branes/orientifolds
-- a type \oprime\ world-sheet.
The coordinate $x^3$ that is being twisted \hetonetwist\ plays, due
to the static gauge condition, a dual role -- it is both a
world-sheet and a spacetime coordinate. 
Theories of this sort have been considered in the past \ortfld\
and play an important role in type I T-duality as well
as in type I -- heterotic string duality \polchwit.

The structure of the theory is compatible with the
$SO(16)\times SO(16)$ type \oprime\ string.
Half of the supersymmetries are broken by the
presence of the boundaries on the world sheet.
The states that live in the bulk of the world-sheet
are eight scalars and their eight superpartners.
On the boundaries of the world-sheet, at $x^3=0,
\pi R_3$ (which describe two orientifold planes),
one finds sixteen fermionic states $F^i$. In this
case they are quantum mechanical degrees of freedom;
the mass shell condition sets the momentum $k$ in 
\feri\ to zero.
Quantizing the theory, the $F^i$ provide discrete
anticommuting labels for states living on the boundaries
of the world-sheet -- they are the $SO(16)\times SO(16)$
Chan-Paton labels. 

It is satisfying to see that 
the gauge groups in type I and heterotic string theories,
whose origins seemed quite 
different in the past, actually stem from the same source.
In both cases the gauge group originates in the fermionic states
$F^i$ in the twisted sector \feri\ which, depending on the choice
of the null current $J$, give rise either to a current algebra
on the world-sheet of a heterotic string or to Chan Paton labels
living on the boundary of a type I world-sheet.

In hindsight, this common source of internal symmetry is not at
all surprising.  
One can understand the appearance of the degrees of freedom $F^i$
on the boundary of the open membrane by passing to the type IIA
string description of figure 1.
Our construction naturally gives us fundamental
two-branes attached to a stack of eight-branes. 
Type IIA string theory also describes these objects, but as solitons.
In that description, one has `$2-8$' open strings in
the theory \ortfld, having one end on the two-brane and the other on
an eight-brane.  These strings are bound to 
the boundary of the two-brane where it attaches to the eight-brane.  
Thus one expects the one-dimensional
boundaries of the open membrane to carry fields with
group indices transforming under $U(n_2)\times U(n_8)$  
\ref\bsv{M. Bershadsky, V. Sadov, and C. Vafa, hep-th/9510225.}.
The orientation projection reduces this symmetry to a combination
of orthogonal/symplectic groups.
These fields are the $F^i$ of \feri.
To see that these degrees of freedom should be identified as
the source of heterotic gauge symmetry, perform
a T-duality transformation on the membrane coordinate $x^3$
transverse to the boundary.  This converts the two-brane to
a one-brane (with worldsheeet $(x^0,x^2)$), 
and turns the eight-brane into a nine-brane
(again in the type IIA string effective description).  The
group indices are now carried by `$1-9$' strings; we find the
solitonic description of the heterotic string in the
type I theory \polchwit.  

\medskip

\noindent{}{\it b) M-theory on $\T^5/\Z_2$.}

In recent developments in string duality, an important 
role was played by type II theories compactified on $K3$
surfaces \refs{\witsdvd,\sduality}. 
Type IIA string theory on $K3$ is dual to the
heterotic string on $\T^4$, which in turn is equivalent
to type I on $\T^4$. When the $K3$ CFT approaches certain
singular points in its moduli space, the type II string
exhibits interesting non-perturbative effects. For example,
in the region of moduli space where a two-cycle on the $K3$
shrinks to zero size (\ie\ an $A_1$ singularity develops)
and certain Kalb-Ramond fields are
suitably adjusted 
\ref\asp{P. Aspinwall, hep-th/9507012.}, 
the type IIA theory
has additional light gauge fields, enhancing
the gauge group to $U(2)$ (from $U(1)\times U(1)$), while a
type IIB string near an $A_1$ singularity has a self-dual string
with tension that goes to zero as one approaches the singularity.
The states becoming light in these two cases can be thought of as arising
from two- and three-branes, respectively, wrapping around the vanishing
cycle. M-theory on a $\T^5/\Z_2$ orientifold provides a useful
picture of all these phenomena, and we will discuss it
in this subsection.
We start with a brief explanation of the role of open
membranes in type IIB theory on $K3$ which will prove useful
for interpreting the results.

It has been shown recently 
\ref\dk{D. Kutasov, hep-th/9512145 (revised version, to appear).}\
that a type II string theory on $\IR^4/\Z_2$, with a certain
$\Z_2$ discrete torsion turned on, describes a vacuum with
an orientifold plane and an NS-NS five-brane sitting at the fixed
point of the orbifold. That result (as well as the related
work of \refs{\bsv,\dm,\witfivebrane}) allows one to study the behavior
of type II theories on $K3$ in a simple manner.

Consider first the type IIA theory on a $K3$ of the
form $(\S^1)^4/\Z_2$. T-duality on one of the circles
turns this into a type IIB string and furthermore turns
on the requisite discrete torsion \dk. The type IIB string thus
lives on a four-torus with sixteen orientifold planes
located at the fixed points of the orbifold, and sixteen
NS-NS  five-branes that are initially located at the
orientifold planes as well\foot{One can use 
the $SL(2, \Z)$ symmetry of type IIB string theory
to turn this type IIB string into one living on a
D-manifold with sixteen orientifolds and sixteen 
Dirichlet five-branes attached to them.}. 
After the T-duality transformation, the blowing-up modes of
the original orbifold that allow one to move in the moduli space
of type IIA on $K3$ turn 
into translational modes that move the five-branes away
from the orientifold planes
(the discrete torsion is essential to this interpretation,
turning degrees of freedom that transform as 
$\underline 3\oplus\underline 1$ of 
the tangent space $SO(4)$
into the $\underline 4$ required for Nambu-Goldstone modes). 
Note that this picture provides a global description
of the $K3$ moduli space.

\vskip 1cm
{\vbox{{\epsfxsize=4in
    \centerline{\epsfbox{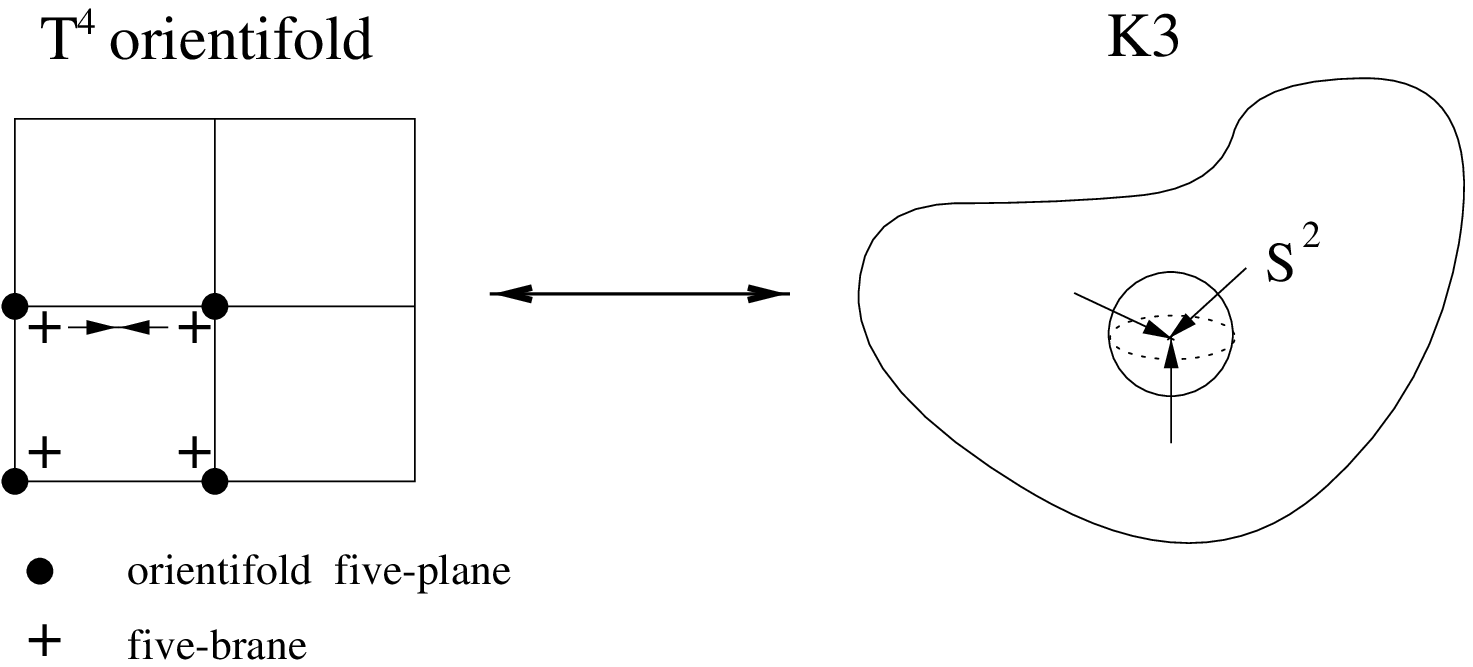}}
\vskip .5cm
    {\raggedright\it \vbox{
{\bf Figure 2.}
{\it Relation between $\T^4/\Z_2$ K3 orbifold and equivalent
configuration of five-branes and orientifold five-planes.
A vanishing two-cycle on the K3 is equivalent to pair of
five-branes approaching one another.}
 }}}}
    \bigskip}

The generic singular limit of the $K3$
CFT corresponds in the T-dual language to a limit in which
two of the five-branes approach one another
(see figure 2). 
When the five-branes are well separated, states corresponding
to strings with one end on each of 
the five-branes 
\foot{These are D-string states if the five-branes
are NS-NS and fundamental string states after the $SL(2, \Z)$
transformation of the previous footnote.} 
are massive. In the limit of coinciding five-branes
their mass goes to zero; these open strings give rise to the additional
light gauge fields that are responsible for the singularity
in the CFT (at least this is the interpretation of
the singularity if one approaches it keeping $g_{\rm string}$ fixed).

A similar story occurs for the type IIB string on $(\S^1)^4/\Z_2$
\refs{\sds,\seibwit}.
T-duality relates it to a type IIA string on a 
four-torus with sixteen orientifold planes corresponding to the fixed
points of the orbifold and sixteen NS-NS five-branes. Again, the
CFT becomes singular when two or more of the five-branes approach 
each other.  This time the objects that can end on the NS-NS fivebranes
are Dirichlet two-branes.  These objects become light
in the limit where the two five-branes coincide, 
transforming into tensionless self-dual 
strings embedded in the five-brane.
One of our purposes in this section
is to study these strings.

An eleven-dimensional interpretation is also available
\refs{\dm,\witfivebrane}, and is closely related to the
above ten-dimensional discussion.  Compactification
of the eleven-dimensional theory on $\T^5/\Z_2$ should be 
equivalent to type IIB theory on a $\T^4/\Z_2$ $K3$ orbifold when
one of the twisted $\S^1$'s shrinks away.
Anomaly cancellation requires 16 five-branes to be distributed
among the 32 fixed points of the orbifold \witfivebrane.
Again, tensionless strings appear in the limit
when two of these five-branes approach one another
\refs{\asselfdual,\townsend}.

In section 3 we discussed the eleven-dimensional supermembrane
vacuum of M-theory by gauging the null vector $J=\partial x^1+\partial
y^1$.  The membrane world-volume was parametrized by $(x^0, x^2, x^3)$.
Here we will construct the state in M-theory on $\T^5/\Z_2$
which corresponds to a stretched open membrane with ends restricted
to five-branes. To do that we split the 2+10 dimensions
of spacetime into two groups of 1+5, and twist one of the two groups.
For example, we can orbifold by the $\Z_2$ twist:
\eqn\fivebrane{\eqalign{
  (x^1,x^3,y^1,y^2,y^3,y^4)&\rightarrow -(x^1,x^3,y^1,y^2,y^3,y^4)\ ;\cr
  (\psi^1,\psi^3,\lam^1,\lam^2,\lam^3,\lam^4)&\rightarrow 
	-(\psi^1,\psi^3,\lam^1,\lam^2,\lam^3,\lam^4)\ ;\cr
	(\psibar^1,\psibar^3)&\rightarrow -(\psibar^1,\psibar^3)\ ;\cr
}}
Note that while the fields $y^1, \cdots, y^4$ are chiral, $x^1$,
$x^3$ are non-chiral; consequently, the orbifold \fivebrane\ is chiral
for the $y^a$ but is a standard, left-right symmetric one for $x^1$ and 
$x^3$.

The twist \fivebrane\  breaks half of the supersymmetries, just as in the
$K3$ compactifications of type II and eleven-dimensional
theories described above.
The orbifold also introduces boundaries both on the
membrane world-volume and in spacetime. To see their role
we now turn to the physical excitation spectrum.

The analysis of the physical spectrum proceeds again in two
stages. In the untwisted sector of the orbifold we find
(compare to \vaa, \gff) eight scalars living in the bulk of the
membrane. Half of them, $(V^2, V^3, V^4, \Phi)$, satisfy Neumann
boundary conditions on the ends of the open membrane; 
their vertex operators have the form \vaa.
The other half
$( V^5, \cdots, V^8)$ satisfy Dirichlet boundary conditions, and
so are fixed at the boundaries; their vertex operators also have the
form \vaa, but with $cos(k_3x^3)$ replaced by $sin(k_3x^3)$
in order to survive the twist \fivebrane.
Note that just as in the previous
subsection, the orientifold $\T^5/\Z_2$ is not directly related
to the fields that appear in \fivebrane. Rather, the $\T^5/\Z_2$
is parametrized by $(V^5, \cdots, V^8)$ and $x^3$ (which serves
as both a membrane world-volume and spacetime coordinate in static gauge).
The world-volume of the five-brane is parametrized by 
$(x^0, x^2, V^2, V^3, V^4, \Phi)$.

The structure in the untwisted sector is similar to that described in
\refs{\asselfdual,\townsend}. 
The membrane with transverse coordinates $V^a$, $\Phi$
has two ends stuck on five-branes. The boundary dynamics 
is described by the six-dimensional self-dual string of \sds.
As we bring the two five-branes closer together, by taking
$R_3\to0$, the string on the boundary becomes tensionless.

The twisted sector of the orientifold doesn't seem to have been
addressed previously. It would describe
states that live on the boundary of the membrane, which
is a string, and could potentially modify the boundary dynamics.
We thus turn next to the twisted sector.

The physical spectrum in the twisted sector
includes the following bosonic states 
\eqn\sdsbos{
W^{i\alpha} =e^{-\phi} \sigma\Sigma^i S^\alpha  
e^{-\phibar_v-\half\phibar_a}\ e^{ik\cdot x}}
where $\sigma$ is a combination of a twist field for
$(x^1, x^3)$, spin fields for $(\psi^1, \psi^3)$ and
$(\bar\psi^1, \bar\psi^3)$, twist fields for the 
left- and right-moving $U(1)$ ghosts, and a twist field
for the left-moving super $U(1)$ ghosts:
\eqn\ssss{\sigma=\sigma_{13}S_{13}\bar S_{13}e^{{\half\tilde\phi}
+{\half\bar{\tilde\phi}}+{\half\rho}}}
$\Sigma^i$ and $S^\alpha$ are chiral twist fields for $(y^1,y^2,y^3,y^4)$
and spin fields for the corresponding fermions. 
The 1+2 dimensional membrane world-volume
$(x^0, x^2, x^3)$ has two 1+1 dimensional boundaries corresponding to
the two fixed points of the $x^3$ orbifold \fivebrane; 
four scalar fields \sdsbos\ (and their four superpartners under the unbroken
supersymmetry) live on each boundary (see the discussion after equation 
\tiid\ for the counting of states). 

What are we to make of the operators $W^{i\alpha}$? 
They are analogs of the sixteen fermionic operators 
$F^i$ \feri\ that we found in our 
$E_8\times E_8$ heterotic construction.
There, they described the $E_8$ current algebra degrees
of freedom living on the boundaries of the open membrane.
We found that the existence of these degrees of 
freedom could be deduced by considering 
(in type IIA language) $2-8$ strings
that are bound to the boundary of the open membrane.
In the current situation, the boundaries of the
open membranes are described in type IIA language 
as residing on Dirichlet four-branes; the
$2-4$ strings that pass between the two-brane and four-brane
are bound to their intersection.
These degrees of freedom generate the
`Chan-Paton fields' $W^{i\alpha}$.
The presence of these fields
on the boundaries of the open membrane clearly
affects the dynamics of the six-dimensional self-dual string
(in particular the latter secretly lives in more than
six dimensions). We will leave a more detailed description of
the self-dual string to future work, 
and conclude this subsection with a few comments
on other theories described by the $\T^5/\Z_2$ orientifold
of M-theory, and an observation about our construction of
self-dual strings.

A T-duality transformation on $(x^3,V^5,\cdots,V^8)$ relates 
the type IIA theory on a $\T^5/\Z_2$ orientifold
to the type I string on $\T^5$; the Dirichlet four-branes become
the nine-branes of the D-brane interpretation of type I
string theory, while the two-brane is converted into a five-brane.
Alternatively,
double-dimensional reduction along $x^2$ 
of the eleven-dimensional description of the membrane 
turns it into an open string, and the five-branes on which it ends into
four-branes. 
The theory again describes a type IIA string on a $\T^5/\Z_2$
orientifold, and again 
is equivalent by T-duality to the type I string
on $\T^5$. 
Dimensional reduction in one of the directions transverse to the
fivebrane (e.g. $V^8$) leads to a type
IIA string state with a Dirichlet two-brane ending on NS-NS fivebranes.
As dicussed in the beginning of this subsection this is a T-dual
description of the sector of the type IIB theory that becomes
massless as the theory approaches an $A_1$ singularity. 
Lastly, consider T-duality on the membrane coordinate $x^3$
transverse to the IIA four-brane; this operation turns 
the four-brane into a Dirichlet five-brane, and the membrane into a
D-string, which is embedded in the five-brane.
In this way, one can make contact with the ideas of
\ref\douglas{M. Douglas, hep-th/9512077.}.

In this subsection, we have described the open membrane of
M-theory through an orientifold which twists half of the
10+2 spacetime coordinates.
The boundary of the membrane is a one-dimensional
object living in a five-brane sitting on the
orientifold plane.
These boundaries are the non-critical strings of \sds.
This self-dual string appears as the boundary dynamics of
an open membrane whose boundary lies in a five-brane
\refs{\asselfdual,\townsend}.
The coincident five-brane limit isolates the boundary dynamics
by freezing the bulk oscillations of the membrane.
One might imagine that another way of isolating the boundary
degrees of freedom is to topologically twist the bulk theory.
This situation is reminiscent of 
the construction of Chern-Simons theory 
from orientifolds in Calabi-Yau manifolds
\ref\csfromstrings{E. Witten, hep-th/9207094.}.  
Consider the sigma model
that describes strings on a six (real) dimensional 
Calabi-Yau manifold.  The A-twisted
topological sigma model describes topological strings
in the Calabi-Yau target.
An orientifold twist on the Calabi-Yau manifold produces
three-dimensional Chern-Simons theory 
on the fixed point set of the involution.
Open strings whose ends rest in the 
fixed point set are described
by this Chern-Simons theory.
In rather precise analogy,
we could begin in $10+2$ dimensions and then
twist $5+1$ of these dimensions to find the self-dual 
string as a membrane boundary lying in the fixed point set of 
the involution.  

This tight parallel between the construction of 
\csfromstrings\ and the orientifold five-brane 
leads to an intriguing speculation.  We expect the 
2+2 dimensional world-volume theory of the M-brane to 
be a supersymmetric theory of self-dual dilaton gravity coupled to 
self-dual matter, with a 10+2 dimensional field space.
It would be interesting to see how much of the additional
structure that appears in the Chern-Simons construction 
might appear here.  Consider a topological twist of this
theory, describing topological M-branes in the target;
combined with a suitable orientifold twist, one expects
to find the self-dual string (or perhaps a topological
version of it) as the theory describing the
dynamics of the M-brane boundary in the half-dimensional
5+1 fixed point set.

%%%%%%%%%%%%%%%%%%%%%%%%%%%%%%%%%%%%%%%%%%%%%%%%%%%%%%%%%%%%%%

\newsec{Comparison of M-brane and D-brane fluctuations}

When the target world-volume theory of the N=2 string is
2+1 dimensional, its field content and supersymmetries 
are compatible with those of the Dirichlet two-brane of
ten-dimensional type IIA strings: seven scalars $V^a$, a vector $A$,
and their eight fermion superpartners.  
However, as noted in \km, in the N=2 string
there is a three-point interaction between two $V$'s and an $A$:
\eqn\threepoint{
\vev{V^a(1)V^b(2)A(3)}=\delta^{ab}(k_1\cdot\bar k_2-k_2\cdot\bar k_1)
	\ \hf\xi\cdot(k_1-k_2)\ ,
}
where $\xi$ is the polarization of the gauge field.
Such a coupling does not appear in the effective action
of the two-brane in trivial backgrounds (it is not even
Lorentz invariant in 2+1 dimensions), so one might wonder
if the proposed identification is correct.

In this section we explore the precise relationship between 
the 2+1 dimensional version of the M-brane and the type IIA
Dirichlet two-brane.  Our main result is that \threepoint\
also arises in the Dirichlet two-brane effective action
in the presence of a constant background world-volume
electromagnetic field (actually the two-brane couples to
the gauge-invariant combination $\FF=dA-B$, where $B$ is the
universal antisymmetric tensor gauge potential of string theory).
The strength and polarization of this background $\FF$ field
are related to the N=2 U(1) current 
$\bar J=\II_{\mu\nu}\psibar^\mu\psibar^\nu$
by $\II_{\mu\nu}\propto\FF_{\mu\nu}$; $\mu,\nu=0,2,3$ 
(maintaining our convention that the left-moving null
current is $J=\d x^1+\alpha_i \d z^i$).

The effective dynamics of the type IIA two-brane is 
governed by the Born-Infeld action
\eqn\borninfeld{
S=\int d^3x\Bigl[e^{-\phi}\sqrt{\det(G+\FF)}
\Bigr]\ .
}
Here $G$ is the induced metric $G_{\mu\nu}=\d_\mu V^a\d_\nu V^a$.
Expand about the static gauge solution
$V^a=\delta_{\mu}^a x^\mu + v^a$ in the presence of a constant
background electromagnetic field 
$\FF_{\mu\nu}=F_{\mu\nu}+\d_{[\mu} a_{\nu]}=F_{\mu\nu}+f_{\mu\nu}$
and dilaton $e^{-\phi}=1/\lambda$:
\eqn\expn{
  \det^{1/2}[G+\FF]=
        \det^{1/2}[G]\ \Bigl[1 + \half G^{\mu\nu}G^{\rho\sigma}
              \FF_{\mu\rho}\FF_{\nu\sigma}\Bigr]^{1/2}}
The background field $F$ breaks 2+1 dimensional Lorentz symmetry
down to the one-dimensional subgroup leaving $F$ fixed.  For instance,
if $F$ is a magnetic field, only spatial rotations remain. 
In such a situation, one generically does not expect a Lorentz-invariant
effective action to result upon expanding \expn\ in powers of
the fluctuations. Remarkably, 
for a background magnetic field 
$F_{\mu\nu} = F\epsilon_{0\mu\nu}$ with 
$\Lambda=[1+\coeff12 F_{\mu\nu}F^{\mu\nu}]^{1/2}$,
the rescaling
$x^0\rightarrow \Lambda x^0$, 
$a_{2,3}\rightarrow\Lambda a_{2,3}$
restores Lorentz invariance of the kinetic terms 
for both $v^a$ and $a_\mu$.  
A similar modification works for other orientations of the background
electromagnetic field.  This rescaling simultaneously produces a 
Lorentz covariant form for the cubic interaction term
\threepoint. To cubic order we have
\eqn\effac{\eqalign{
S=\frac1\lam\int d^3x \Bigl[\Lambda^2 &+ \hf \d v^a\d v^a
	+ \coeff14 f^{\mu\nu}f_{\mu\nu} \cr
  & - \Lambda^{-1} \d^\mu v^a\d^\nu v^a\bigl(F_{\mu\rho}f_\nu{}^\rho -
    \coeff14 \eta_{\mu\nu}F_{\lam\rho}f^{\lam\rho}\bigr) 
       +\ldots\Bigr]\ ,
}}
in perfect agreement with the N=2 string result \threepoint\
provided $F_{\mu\nu}/(1+\half F^2)^{1/2}=\II_{\mu\nu}$. This relation
fixes the background electromagnetic field to a value 
determined directly by the choice of N=2 $U(1)$ current.
In other words, the background electromagnetic field breaks
$O(1,2)$ symmetry on the D-brane down to $O(2)$ or $O(1,1)$ 
in precisely the same way that the N=2 U(1) current breaks
$O(2,2)$ symmetry on the M-brane down to $U(1,1)$ or $GL(2,\IR)$.
Since the normalization of the U(1) current is fixed by
the central charge of the N=2 algebra, we see once again that
the N=(2,1) string produces a very specific background
of M-theory.

%%%%%%%%%%%%%%%%%%%%%%%%%%%%%%%%%%%%%%%%%%%%%%%%%%%%%%%%%%%%%%

\newsec{Discussion}

\subsec{Summary.}

Using the definition of M-theory proposed in \km,
we have initiated the detailed 
analysis of its vacua.
We have discussed three broad classes of vacua.
In section three, we showed that type IIA/B string and
eleven-dimensional supermembrane theories are different
limits within a single moduli space.
In section four, we presented two additional classes,
corresponding to M-theory on $\S^1/\Z_2$ and $\T^5/\Z_2$
orientifolds.  The first of these describes type I and
heterotic string theories; the second yields type II strings
on K3, type I strings on $\T^4$, and the six-dimensional
self-dual string.

These constructions provide valuable insight into the
nature of M-theory.  Among the important lessons are:

\item{i)}
Our results give strong new evidence for the validity of
many conjectures regarding the properties of M-theory,
its relation to different string vacua, and the various
(S-, T-, and U-) dualities that relate them.
Conversely, this evidence strongly supports the proposal of \km\
for the dynamical structure underlying M-theory.

\item{ii)}
The unified picture of string and membrane
world-volume theories treats them
as different reductions of 2+2 dimensional self-dual
dynamics.  This self-duality structure is imposed by the right-moving
N=2 dynamics of the (2,1) string.
The spacetime in which these 2+2 `M-brane'
world-volumes is embedded is 10+2 dimensional.
This spacetime information is carried by the left-moving 
N=1 part of the (2,1) string, which typically has eight 
physical massless
bosonic excitations together with fermionic partners
related by Green-Schwarz type supersymmetry.
The 2+2 dimensions of the world-volume together with
these eight bosonic Nambu-Goldstone modes
show that the world-volume is embedded in 10+2 dimensions. 

\item{iii)}
Our formalism treats the modes on the M-brane
in a unified fashion as the dimensional reduction of 
a twelve-dimensional gauge field.  
Reminiscent of the D-brane construction \polch,
the M-brane splits these into a gauge field on the brane
(the Kahler vector potential) and a set of scalars
(the transverse fluctuations of the brane).
Higher-dimensional branes make their appearance in
mixed momentum/winding sectors of the N=2 string, when all
spatial directions are compactified.

\item{iv)}
Fundamental gauge symmetry in M-theory has a single
source -- boundaries of open membranes.
Until now, the gauge group of the type I theory and that of
the heterotic string seemed to come from completely different
sources (although they were connected through a chain of dualities).

\item{v)}
String worldsheet dynamics is governed by conformal field theory;
a beautiful generalization of this structure is emerging.
The 2+2 dimensional M-brane dynamics typically describes membrane
world-volumes; string worldsheets arise in particular limits.
We have seen that many structures of conformal field theory, such as 
worldsheet/spacetime orbifolds, carry over to this new setting.

\noindent
Thus our results provide an explicit construction of
strings and membranes in a unified fashion, and demonstrate
conclusively that the appropriate setting for M-theory
is twelve-dimensional.
Many conjectures about M-theory
are placed on a solid footing.
We have presented a small portion of the rich variety
of constructions that are available within our framework.
We expect that further elaborations will shed yet more light
on the structure of M-theory.

\subsec{Directions for future work.}

Our results strongly suggest that 
a 2+2 dimensional world-volume theory of membranes is
the appropriate setting for M-theory.
They furthermore suggest that this theory is quantizable.
There exist other formulations of membrane dynamics --
the Born-Infeld and Howe-Tucker
\ref\howet{P. Howe and R. Tucker, \jpa{10}{1977}{L155}.}
actions -- that do not lead to a quantizable theory.
In this sense, the 2+2 dimensional theory should play
in M-theory the role that the Polyakov formulation does
in string dynamics, as compared to the Nambu-Goto 
formulation\foot{Indeed, we have seen in section 5 that 
(to the order we have checked)
this self-dual dynamics is classically equivalent to that
of the Born-Infeld action for membranes.}.

A direct understanding of the 2+2 dimensional world-volume
theory would be desirable.  To achieve that, we
propose to extract information about such a theory
from the N=(2,1) string, whose target space is
this world-volume.
The N=(2,1) sigma-model was analyzed in \km\ precisely to
begin addressing this issue.  
In any string theory, the sigma
model tells us about the target space geometry, and gives
off-shell information (for instance about gauge invariances
and field content).  Thus the N=(2,1) sigma-model geometry
is precisely the classical geometry of the 2+2 dimensional M-brane
world-volume theory.
The structure that emerged from this
analysis was a theory of self-dual dilaton gravity
coupled to self-dual matter (\eg\ self-dual $[U(1)]^{12}$
or $[U(1)]^{28}$ gauge theory).  
The symmetries of self-dual geometry play the same role
here that conformal invariance does in string theory.
Moreover, these symmetries lead to the integrability
of the classical field equations.

The existence of (2,1) supersymmetry
imposes constraints on the sigma-model background. 
The N=2 right-moving supersymmetry requires the geometry
to be self-dual.
The N=1 left-moving supersymmetry
solders four of the fiber
directions to the tangent bundle of the M-brane
(\ie\ embeds the spin connection in the gauge group).
Indirectly, this then describes the embedding of the 
M-brane into spacetime.
In addition, the left-moving null current algebra
of the N=(2,1) string construction of the M-brane
requires the existence of a covariantly constant null Killing
vector along the fibers, compatible with the constraints
that lead to (2,1) supersymmetry.  
The orientation of this null Killing vector relative
to the embedded M-brane tangent bundle 
determines whether one finds a string
or a membrane. 

In other situations with analogous structures of symmetries
and integrability, one is able to use Ward identities
and/or geometric ideas to quantize the theory.
While we have no proof, we strongly suspect that 2+2 dimensional
self-dual dilaton gravity coupled to matter exists as a quantum
theory.  The theory should be
strongly constrained by the Ward identities arising from the
world-volume current algebra.  Quantization of four-dimensional
self-dual matter in this spirit has been discussed in
\ref\lmns{A. Losev, G. Moore, N. Nekrasov, and S. Shatashvili,
hep-th/9509151.}.

We conclude from this discussion
that it is important to understand the
quantization of self-dual dilaton gravity coupled to matter
in 2+2 dimensions; to construct its algebra of
Ward identities along the lines of
\ref\bpz{A.A. Belavin, A.M. Polyakov, and A.B. Zamolodchikov,
\np{241}{1984}{333}.}; and to understand the null reduction
that gives strings and membranes.

Finally, there are (at least) two possible attitudes as to what 
basic objects underly M-theory.  First, one might suppose
that there is a lowest-dimensional object -- a string or
membrane -- which plays a fundamental role, such that
the other $p$-branes appear as collective excitations.
A second viewpoint posits that all $p$-branes are
of equal importance \ref\demo{P. Townsend, hep-th/9507048.}.
A striking feature of the N=(2,1) string construction
is the unification of all the dilaton-gravity-matter
fields on the M-brane into what looks like a 
twelve-dimensional gauge field (since they all come from the
left-moving polarization states of the N=1 superstring).
As in the D-brane construction \polch,
the M-brane splits these into a gauge field on the brane
and a set of scalars; but the
spatial dimensions of the M-brane are at most two.
This leads one to wonder whether there might be a generalization
of the M-brane world-volume theory that leads to a consistent
quantization of all $p$-branes, perhaps involving 10+2
dimensional Yang-Mills theory.  
Hints of this are found in the mixed momentum/winding sectors
of the (2,1) string when all spatial dimensions are compact.
Could the restriction to 
strings and membranes be yet another artificial restriction
of the N=(2,1) string that disappears in the final
formulation of M-theory?

%%%%%%%%%%%%%%%%%%%%%%%%%%%%%%%%%%%%%%%%%%%%%%%%%%%%%%%%%%%%%%

\vskip 1in
{\sl Acknowledgements}:
It is a pleasure to thank 
J. Harvey
for discussions.

\listrefs
\bye